\documentclass[12pt,a4paper,final]{article}
\usepackage[latin9]{inputenc}
\usepackage{times}
\usepackage{slashed}
\usepackage{a4wide}
\usepackage{amssymb}
\usepackage{amsmath}
\usepackage{bbm}
\usepackage{dsfont} 
\usepackage{pifont} 
\usepackage[notref,notcite]{showkeys} 
\usepackage{ifpdf}
\ifpdf
\usepackage[pdftex]{graphicx}
\usepackage[pdftex,unicode,implicit]{hyperref}
\hypersetup{%
  pdftitle    = {Ultracold spherical horizons in gauged N=1,d=4 supergravity},
  pdfkeywords = {supersymmetry, black holes, De Sitter},
  pdfauthor   = {P. Meessen and T. Ortin},
  pdfcreator  = {pdf\LaTeXe\ with package \flqq hyperref\frqq},
  pdfproducer = {pdf\LaTeXe\ with package \flqq hyperref\frqq},
  pdfpagemode = None,  
  pdffitwindow= true,  
  unicode     = true,
  plainpages  = false,
  colorlinks  = true,  
  citecolor   = blue,
  urlcolor    = red,
  linkcolor   = black
}
\newcommand{\hepth}[1]{{\tt \href{http://www.arXiv.org/abs/hep-th/#1}{hep-th/#1}}}
\newcommand{\hepph}[1]{{\tt \href{http://www.arXiv.org/abs/hep-ph/#1}{hep-ph/#1}}}
\newcommand{\grqc}[1]{{\tt \href{http://www.arXiv.org/abs/gr-qc/#1}{gr-qc/#1}}}

\newcommand{\arxiv}[1]{{\tt \href{http://www.arXiv.org/abs/#1}{arXiv:#1}}}
\else
  \usepackage[dvips]{graphicx}
  \usepackage[unicode,implicit]{hyperref}
  \newcommand{\hepth}[1]{{\tt hep-th/#1}}
  \newcommand{\hepph}[1]{{\tt hep-ph/#1}}
  \newcommand{\grqc}[1]{{\tt gr-qc/#1}}
  
  \newcommand{\arxiv}[1]{{\tt arXiv:#1}}
\fi

\begin{document}
\begin{flushright}
\small
FPAUO-10/04\\
IFT-UAM/CSIC-10-45\\
{\bf arXiv:yymm.nnnn}\\
July $22^{\rm nd}$, $2010$
\normalsize
\end{flushright}
\begin{center}
\vspace{2cm}
{\Large {\bf Ultracold Spherical Horizons}}\\[.5cm]
{\Large {\bf in Gauged $N=1$, $d=4$ Supergravity}}
\vspace{2cm}

 {\sl\large Patrick Meessen}
 \footnote{E-mail: {\tt meessenpatrick@uniovi.es}}$^{\dagger}$,
{\sl\large and Tom{\'a}s Ort\'{\i}n}
\footnote{E-mail: {\tt Tomas.Ortin@cern.ch}}$^{\ddagger}$

\vspace{.8cm}

$^{\dagger}${\it Department of Physics, University of Oviedo,\\
Avda. Calvo Sotelo s/n, E-33007 Oviedo, Spain}

\vspace{.3cm}

$^{\ddagger}${\it Instituto de F\'{\i}sica Te\'orica UAM/CSIC\\
Facultad de Ciencias C-XVI,  C.U.~Cantoblanco,  E-28049 Madrid, Spain}\\

\vspace{2cm}


{\bf Abstract}

\end{center}

\begin{quotation}\small
  We show that the near-horizon limit of ultracold magnetic
  Reissner-Nordstr\"om-De Sitter black holes, whose geometry is the direct
  product of 2-dimensional Minkowski spacetime and a 2-sphere, preserves half
  of the supersymmetries of minimal R-gauged $N=1$, $d=4$ supergravity.
\end{quotation}

\newpage

\pagestyle{plain}




The supersymmetric black-hole solutions of 4-dimensional supergravity theories
can often be understood as solitons interpolating between two maximally
supersymmetric vacua of the theory to which they approach in the far-field and
near-horizon (NH) regions.  The NH geometry (the product space $AdS_{2}\times
S^{2}$, known as Bertotti-Robinson solution, in the typical
asymptotically-flat cases) contains a great deal of information about the
constituents of the original solution and is amenable to a dual description by
a gauge theory living in the boundary of the $AdS_{2}$ space. In particular,
the radius of the $S^{2}$ factor of the NH geometry, which corresponds to a
horizon with the same topology, is directly related to the entropy. The
sufficiency of the NH description to describe the black-hole entropy,
independently of the asymptotic behavior of the solution, is a consequence of
the attractor mechanism \cite{Ferrara:1995ih}.

The topology of the spatial factor of the NH solution ($S^{2}$ in the above
example) coincides with that of the spatial sections of the black-hole
horizon. In the 4-dimensional, asymptotically flat (vanishing cosmological
constant) case, a classical theorem by Hawking
\cite{Hawking:1971vc}\footnote{A recent paper with references on the
  generalization of this theorem to higher dimensions and non-vanishing
  cosmological constant is \cite{Racz:2008tf}.}. and the ``topological
censorship theorem'' of Ref.~\cite{Friedman:1993ty} constrain that topology
to be that of $S^{2}$. However, black holes with event horizons topologically
inequivalent to $S^{2}$ have been discovered in dimensions higher than four
\cite{Emparan:2001wn,Elvang:2004rt}\footnote{For a recent review with
  references, see \cite{Emparan:2008eg}.}; in four dimensions and in presence of a negative
cosmological constant \textit{topological black holes} with horizons which are
Riemann surfaces or arbitrary genus have also been constructed
\cite{Lemos:1994xp}.

The requirement of unbroken supersymmetry of the NH solution strongly
constrains the possible NH geometries and horizon topologies. In
Ref.~\cite{Reall:2002bh} Reall showed that, in 5-dimensional supergravity,
supersymmetry only allows three possible horizon topologies: $T^{3}$,
$S^{1}\times S^{2}$ (the topology of the supersymmetric black ring of
Ref.~\cite{Elvang:2004rt}) and (possibly a quotient of) a homogeneously
squashed $S^{3}$. On the other hand, in Ref.~\cite{AlonsoAlberca:2000cs} it
was shown that only the genus bigger than one horizons may have unbroken
supersymmetry in minimal gauged $N=2$ $d=4$ supergravity.

In a recent paper \cite{Gutowski:2010gv} Gutowski and Papadopoulos have
studied possible topologies of supersymmetric horizons of black hole solutions
of $N=1,d=4$ supergravity finding that, regardless of the supersymmetry
properties of the complete black-hole solution\footnote{There are no
  supersymmetric asymptotically flat black-hole solutions in $N=1,d=4$
  supergravities \cite{Gran:2008vx,Ortin:2008wj}. No supersymmetric black
  holes with other asymptotic behaviors are known, either.}, if the horizon is
compact and supersymmetric (i.e.~if the NH geometry is), then its constant
time sections have to be, topologically, tori. Our purpose in this note is to
investigate possible simple realizations of these NH geometries in simple
$N=1,d=4$ supergravity theories.

Since, in order to have topological black hole solutions, we need a negative
cosmological constant, we should consider a $N=1,d=4$ theory providing a
minimal supersymmetric embedding of the cosmological Einstein-Maxwell
(EM-$\Lambda$) theory

\begin{equation}
S= \int dx^{4}\sqrt{|g|}\, \{ R   -F^{2} - \Lambda\}\, ,
\end{equation}

\noindent
for negative (aDS) cosmological constant $\Lambda$.  However, just to be
general (which will prove fortunate in he end), we are also going to consider
another $N=1,d=4$ supergravity theory providing a supersymmetric embedding of
EM-$\Lambda$ for positive (DS) cosmological constant.

The bosonic equations of motion of these two theories take the common form

\begin{eqnarray}
R_{\mu\nu} & = & \frac{\Lambda}{2} g_{\mu\nu} 
         \ +\ 
2\left[ F_{\mu}{}^{\rho}F_{\nu\rho} -\textstyle{1\over 4}\ g_{\mu\nu}F^{2} \right] \; ,\\
& & \nonumber \\
d\star F & = & 0 \; .
\end{eqnarray}

\noindent
The two theories that we are going to consider have the same matter content:
an Abelian vector supermultiplet $\{A_{\mu},\lambda\}$ coupled to the
$N=1,d=4$ supergravity multiplet $\{e^{a}{}_{\mu},\psi_{\mu}\}$. The first
theory, constructed by Townsend in Ref.~\cite{Townsend:1977qa}, has, in more
modern parlance, a constant superpotential $W=g/2$ which gives a negative
cosmological constant $\Lambda= -8g^{2}$ and possesses a maximally
supersymmetric $aDS_{4}$ solution.  In the second theory, constructed by
Freedman in Ref.~\cite{Freedman:1976uk}, the Abelian vector field is used to
gauge the global $U(1)$ R-symmetry via a Fayet-Iliopoulos term which gives a
positive cosmological constant $\Lambda= +g^{2}/2$ where $g$ is the gauge
coupling constant. These two possibilities cannot be combined because the
constant superpotential breaks R-symmetry.

Although the bosonic sectors of these two theories are identical, up to the
sign of the cosmological constant, the couplings of the fermionic sectors and
the supersymmetry transformations are substantially different, which results in
very different supersymmetric configurations even though all the Killing
spinors of the supersymmetric configurations of any $N=1,d=4$ supergravity
must satisfy the condition \cite{Gran:2008vx,Ortin:2008wj}

\begin{equation}
\gamma^{u}\epsilon=0\, ,  
\end{equation}

\noindent
where $u$ is a null coordinate, or, equivalently

\begin{equation}
\gamma^{01}\epsilon = \pm\epsilon\, .
\end{equation}


\section{Supersymmetry of Pleba\'nski-Hacyan geometries}
\label{sec-susy}

We are going to consider configurations whose metric is the direct product of
two 2-dimensional subspaces of constant curvature, the first one parametrized
by the first two (timelike and spacelike) coordinates and the second one
parametrized by the last two (spacelike) coordinates.  This generic class of
solutions to EM-$\Lambda$ was first obtained by Pleba\'nski {\&} Hacyan in
Ref.~\cite{art:PlebHacyan1979}, and includes as special cases the
Bertotti-Robinson solution ($aDS_{2}\times S^{2}$) and the Nariai universe
($DS_{2}\times S^{2}$) \cite{art:nariai}, whose discovery predates
the work \cite{art:PlebHacyan1979}.

The geometry of the purely spacelike 2-dimensional subspace is expected to
correspond to that of the constant-time sections of a black-hole horizon. The
Maxwell field will have non-vanishing components $F_{01}=\alpha$ and
$F_{23}=\beta$, where $\alpha$ and $\beta$ are real constants (that is: the
components of the Maxwell field are proportional to the volume 2-forms of the
two subspaces). We will make this more precise Ansatz later on.


\subsection{$N=1,d=4$ Supergravity with constant superpotential}
\label{sec-susy1}

As was mentioned before, the minimal version of this theory was constructed by Townsend in
Ref.~\cite{Townsend:1977qa} and when coupled to a vector multiplet corresponds to a supersymmetric version of the
EM-$\Lambda$ theory with the cosmological constant $\Lambda = -8g^{2}$ being
of the anti-De Sitter kind.  The supersymmetry transformations of the fermions
for vanishing fermions are\footnote{
  For clarity's sake we mention that we are using a normalized version of the slash, 
  {\em i.e.\/} for the 2-form $F$ we have $2\slashed{F} \equiv F_{ab}\gamma^{ab}$.
}

\begin{eqnarray}
\label{eq:Town1}
0\; =\; \delta_{\epsilon}\psi_{\mu} & = & 
\nabla_{\mu}\epsilon 
+\tfrac{i}{2}g\gamma_{\mu}\epsilon^{*}\, ,\\
& & \nonumber \\
\label{eq:Town2}
0\; =\; 2\delta_{\epsilon}\lambda & = &
\not\! F^{+}\epsilon\, ,
\end{eqnarray}

\noindent
where $\nabla$ is the general and Lorentz-covariant derivative.


That this theory does not admit supersymmetric solution of the type we are
after is easily deduced by calculating the integrability condition for
Eq.~(\ref{eq:Town1}):

\begin{equation}
\left[\ \slashed{R}_{\mu\nu} \ +\ g^{2}\gamma_{\mu\nu}\ \right]\epsilon \; =\; 0\, .   
\end{equation}

\noindent
The split into 2-dimensional spaces of constant curvature, implies that {\em
  e.g.\/} $\slashed{R}_{02}=0$, which immediately implies that $\epsilon =0$,
whence no supersymmetric PH solutions exist.


\subsection{Minimal gauged $N=1,d=4$ supergravity}
\label{sec-susy2}

This $N=1$ $d=4$ theory was constructed by Freedman in
Ref.~\cite{Freedman:1976uk} and has the curiosity that it corresponds to
supergravity theory with a De Sitter-like cosmological constant ($\Lambda =
g^{2}/2$).  The relevant supersymmetry transformations for vanishing fermions
are

\begin{eqnarray}
\label{eq:Freed1}
0\; =\; \delta_{\epsilon}\psi_{\mu} 
& = & 
\left[\nabla_{\mu} +\tfrac{i}{2}gA_{\mu}\right]\epsilon \, ,
\\
& & \nonumber \\
\label{eq:Freed2}
0\; =\; \delta_{\epsilon}\lambda 
& = &
\left[ \slashed{F}^{+}\, -\, \textstyle{i\over 2}\ g\ \right]\epsilon\, .
\end{eqnarray}

De Sitter spacetime is a solution of the theory but breaks all supersymmetries.

The Killing spinor equation (\ref{eq:Freed2}) only admits solutions for our
Ansatz if $\alpha=0$ and

\begin{equation}
\label{eq:Freed3}
 \beta \ =\ \pm g/2 
\hspace{.5cm}\mbox{and}\hspace{.3cm}
\left[\ 1\ \pm\ i\gamma^{23}\ \right]\epsilon \ =\ 0\, ,
\end{equation}

\noindent
so that we are dealing with a purely magnetic configuration. 
\par
The integrability condition of the Killing spinor equation (\ref{eq:Freed1}) reads

\begin{equation}
\left[\ \slashed{R}_{\mu\nu} \, -\, ig\ F_{\mu\nu}\ \right]\epsilon \; =\; 0\, .   
\end{equation}

\noindent
The product structure of the metric that we have assumed indicates that the
first factor must be flat 2-dimensional Minkowski spacetime and the second a
2-sphere whose curvature is related to $\beta$ and, therefore, to $g$. 

At this point a more precise form for the Ansatz becomes necessary: using
standard spherical coordinates for the 2-sphere we write

\begin{equation}
\begin{array}{rcl}
ds^{2} & = & dt^{2} - dx^{2} - R^{2}(d\theta^{2} +\sin^{2}\theta d\phi^{2})\,
,
\\
& & \\
A_{\phi} & = & -\beta\ R^{2}\ \cos\theta \, .
\end{array}
\end{equation}

\noindent
The non-vanishing components of the Ricci and Maxwell field strength tensors
are, in the obvious tetrad basis

\begin{equation}
R_{22}\ =\ R_{33}\ =\ -\frac{1}{R^{2}}\, ,
\hspace{1cm}
F_{23} \ =\ \beta\, .  
\end{equation}

The Maxwell equations are automatically solved as the field strength is an
invariant 2-form on a symmetric space; the Einstein equations are solved if

\begin{equation}
R^{2} \; =\; \frac{g^{2}}{4} \ +\ \beta^{2}\, , 
\end{equation}

\noindent
which due to Eq.~(\ref{eq:Freed3}) implies:

\begin{equation}
\label{eq:Freed10}
R \; =\; \sqrt{2}/g \; .
\end{equation}

In order to finish the analysis we need to solve the Killing spinor equations
(\ref{eq:Freed1}); the $0$, $1$, $2$ components are trivial and are solved for
any $t$, $x$ and $\theta$ independent spinor. The last component is also
trivially satisfied once we take into account the following relation
between the spin and the gauge connections $A_{3}= \pm\ g^{-1}\ \omega_{323}$ and
use the projection in Eq.~(\ref{eq:Freed3}).

In conclusion we found a half-BPS solution to Freedman's gauged $N=1$ $d=4$
supergravity that is purely magnetic and whose geometry is
$\mathbb{R}^{1,1}\times S^{2}$. The obvious question then is: can this geometry
be the NH limit of a black hole?  A first naive worrisome point is about the
occurrence of the $\mathbb{R}^{1,1}$ factor in the NH geometry, as the usual
one of supersymmetric black holes would not give rise to $\mathbb{R}^{1,1}$
but rather to $aDS_{2}$. But, as said, this is a naive preoccupation as,
following Gutowski {\&} Papadopoulos, we are asking for the NH-geometry to be
supersymmetric and not the complete solution. If we then couple this to the
fact that the NH geometry of black holes with non-vanishing temperature, such
a Schwarzschild's, leads to a 2-dimensional Rindler space which is locally
isometric to $\mathbb{R}^{1,1}$, the preoccupation should cease to exist. So
in order to find the candidate black hole whose NH-limit gives rise to the
supersymmetric solution, we should analyze the NH-limits of
magnetically-charged black holes with spherical topology in De Sitter spaces.


\section{Reissner-Nordstr\"om-De Sitter black holes}
\label{sec:RNDSbhs}

\begin{figure}
  \centering
  \includegraphics[height=4cm]{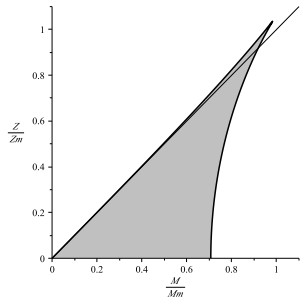}
  \caption{
     A plot of the values of $M$ and $Z$ for which the RNDS
     black holes exist. The straight line are the extreme bh's, {\em
       i.e.\/} the ones for which $M^{2}=Z^{2}$.
  }
  \label{fig:DSbhSpectrum}
\end{figure}

The Reissner-Nordstr\"om-De Sitter (RNDS) black holes can be written in
standard coordinates as

\begin{eqnarray}
\label{eq:31}
ds^{2} & =& fdt^{2} \ -\ f^{-1}dr^{2} \ -\ r^{2}dS^{2}_{[\theta ,\varphi ]}
\; ,\\
& & \nonumber \\
A & =& \frac{Q}{r}\ dt \; -\; P\cos\theta d\varphi \; ,
\end{eqnarray}

\noindent
where $dS^{2}_{[\theta ,\varphi ]} $ stands for the round metric on $S^{2}$
with coordinates $\theta$ and $\varphi$, and the function $f=f(r)$ is given by

\begin{equation}
  \label{eq:1}
  f \; =\; -\frac{\Lambda}{6}r^{2} \ +1\ -\frac{2M}{r} \ +\
  \frac{Z^{2}}{r^{2}}\, ,
 \hspace{.4cm}\mbox{with}\; 
  Z^{2} \ \equiv\ Q^{2}+P^{2} \; .
\end{equation}

As is well-known De Sitter black holes need not exist for all values of the
mass, $M$, and the electro-magnetic charge, $Z$; a plot of the pairs $(M,Z)$
that can give rise to black holes are indicated in
Fig.~(\ref{fig:DSbhSpectrum}) by the grey area and its boundary.  As is
paramount from the figure $M$ and $|Z|$ are bounded by maximal values that in
our normalization of $\Lambda$ are given by

\begin{equation}
  \label{eq:2}
  M_{crit}\; =\; \frac{2}{3\sqrt{\Lambda}} \; =\; \frac{2\sqrt{2}}{3g}
  \hspace{.5cm}\mbox{and}\hspace{.5cm}
  Z^{2}_{crit} \; =\; \frac{1}{2\Lambda} \; =\; \frac{1}{g^{2}} \; .
\end{equation}

\noindent
A point in the grey area corresponds to a black hole with three horizons,
namely an inner one at $r=r_{i}$, an outer one at $r=r_{o}$ and a cosmological
horizon at $r=r_{c}$, the nomenclature deriving from the fact that
$0<r_{i}<r_{o}<r_{c}$. Furthermore, all these horizons are {\em warm} in the
sense that they correspond to single zeroes of $f$, whence one can associate a
temperature to at least the outer and the cosmological horizon.\footnote{ As
  is well-known by expanding $f$ in Eq.~(\ref{eq:1}) around the horizon
  location $r=r_{H}$ as $f=(r-r_{H})\ h(r)$ with $h$ being regular at $r_{H}$,
  one finds that the NH geometry is that of a Rindler space of temperature $T=
  h(r_{H})/(4\pi )$ times a 2-sphere of radius $r_{H}$.  }

The left boundary corresponds to those black holes for which the inner and the
outer horizon coincide $0<r_{i}=r_{o}<r_{c}$, implying that this coincident
horizon, but not the cosmological horizon, has zero temperature: these black
holes are called {\em cold black holes}.  The right boundary corresponds to
the situation where the outer and the cosmological horizons coincide
$0<r_{i}<r_{o}=r_{c}$ and are also cold black holes; they receive the name
{\em Nariai} black holes.  The intersection of these two boundaries,
corresponding to the pair $(M_{crit},Z_{crit})$, for which all three horizons
coincide, goes by the name {\em ultracold black hole} \cite{Romans:1991nq}.

This small discussion then brings us to the question: How are we to identify
the RNDS black-hole solution whose NH limit gives us the supersymmetric
Pleba\'nski-Hacyan solution? The answer is simple: by looking at the NH limit
of the gauge field!  First of all, a non-zero $Q$ would lead to a non-zero
$F_{01}$ so we will take $Q=0$. The NH limit of the vector field strength for
a horizon located at $r=r_{H}$ is 

\begin{equation}
\label{eq:3}
F \ =\ d( -P\cos\theta \ d\varphi ) \ =\ P\ d\theta \wedge\ \sin\, \theta d\varphi 
\ \longrightarrow\ \frac{P}{r_{H}^{2}}\ e^{2}\wedge e^{3} \; ,
\end{equation}

\noindent
and leads to the identification that $P=\beta \ r_{H}^{2}$. Seeing that the
value of $\beta$ for the supersymmetric solution is given in
Eq.~(\ref{eq:Freed3}) and that $r_{H}$ is effectively the radius of the
2-sphere in the NH limit, Eq.~(\ref{eq:Freed10}), we can deduce that our
candidate black hole must have

\begin{equation}
\label{eq:6}
P\; =\; \beta\ r_{H}^{2} \; =\; \pm \frac{g}{2}\ 
\left( \frac{\sqrt{2}}{g}\right)^{2} \; =\; \pm 1/g \; ,
\end{equation}

\noindent
implying that our candidate black hole is none other than the ultracold black hole.

This poses, however, an immediate problem, one already pointed out by Romans
\cite{Romans:1991nq}: as the horizon of the ultracold black hole corresponds
to a triple zero of the function $f$ in Eq.~(\ref{eq:1}), the naive NH limit
does not give as NH geometry Rindler space times $S^{2}$ but a different one,
one that is not even a solution to the equations of motion: the reason for this is that in
this case the usual procedure of zooming in does not conform to Geroch's
criteria of limiting spaces \cite{Geroch:1969ca}.

There is an alternative limiting procedure that does give rise to the desired
result \cite{Ginsparg:1982rs,Cardoso:2004uz} which basically consists in going
first to the cold limit in which $f(r)$ has a double zero and then taking
the NH limit simultaneously with the ultracold limit in a particular way. The
result is the supersymmetric Pleba\'nski-Hacyan solution\footnote{Notice that
  we can arrive at the same result in a more pedestrian way by taking the NH
  limit of a warm or a cold horizon in a first step and then taking the
  ultracold limit in a second step. In the first case, we arrive at the NH
  geometry Rindler$_{2}\times S^{2}$ in the first step and then adjust the
  physical parameters to those of the supersymmetric PH solution in the
  second. In the second case, we arrive to the NH geometry $aDS_{2}\times
  S^{2}$ in the first step while the second step flattens out the $aDS_{2}$
  factor because the ultracold limit is the limit of infinite $aDS$ radius. We
  get the same result in all cases.} which can, therefore, be identified as
the NH limit of the ultracold, purely magnetic, RNDS black hole.

\section{Conclusions}
\label{sec-conclusions}

In this letter we have tried to find simple examples of supersymmetric
horizons in $N=1,d=4$ supergravity theories motivated by the prediction made
in Ref.~\cite{Gutowski:2010gv} that, if any, their spatial sections would
always be topologically equivalent to tori. We have focused on two $N=1,d=4$
theories (Freedman's and Townsend's) whose bosonic sector is the cosmological
Einstein-Maxwell theory with positive and negative cosmological constant,
respectively, and on candidate near-horizon geometries which are the direct
product of two 2-dimensional spaces of constant curvature. We have shown that
none of our candidates is supersymmetric in Townsend's theory ($\Lambda <0$)
but we have also shown that one of them, with the geometry
Minkowski$_{2}\times S^{2}$ is actually supersymmetric in Freedman's ($\Lambda
>0$). Then we have shown that this supersymmetric solution is the NH limit of
the ultracold RNDS black-hole solution when the NH limit is correctly
computed, which means that, even though no RNDS black-hole solution is
supersymmetric, the horizon of the ultracold one, which has the topology of
$S^{2}$, is. We can also say that the non-supersymmetric ultracold RNDS black
hole solution interpolates between non-supersymmetric DS spacetime at infinity
and a half-supersymmetric Pleba\'nski-Hacyan solution at the horizon.

This result is a clear counterexample for the generic prediction of
Ref.~\cite{Gutowski:2010gv}. The reason why our spherically-symmetric NH
geometry was missed is, as far as we can see, that the analysis made in that reference is based on a
gravitino Killing spinor equation that is not general enough, and in particular does not
include Freedman's theory.

Of course, our results do not imply that these are the only possible
supersymmetric NH geometries nor that Freedman's theory and its generalizations
are the only possible $N=1,d=4$ supergravities in which supersymmetric NH
geometries can be found.

At this moment we do not have a clear physical interpretation of this
result. We can only stress the fact that the supersymmetric solution has mass
and magnetic charge which are extremized for a given value of the
cosmological/coupling constant. Furthermore, we would like to point out that,
while Townsend's theory is sometimes called $N=1,d=4,aDS$ supergravity,
Freedman's (studied, for instance, in
Refs.~\cite{Chamseddine:1995gb,Castano:1995ci}) is very different from a naive
(and inconsistent) $N=1,d=4,DS$ supergravity and can be embedded in string
theory \cite{Cvetic:2004km}.

As a final comment let us point out that a fake version of Freedman's gauged
supergravity can be constructed and the existence of fake-supersymmetric
NH-geometries can be studied, which shows that indeed there is a
fake-supersymmetric $aDS_{2}\times \Sigma_{g>1}^{2}$ solution. One can then
also show that there is no $aDS$-black hole which has this NH-geometry.


\section*{Acknowledgments}

This work has been supported in part by the Spanish Ministry of Science and
Education grants FPA2006-00783 and FPA2009-07692, a Ram\'on y Cajal fellowship
RYC-2009-05014, the Comunidad de Madrid grant HEPHACOS S2009ESP-1473, the
Princip\'au d'Asturies grant IB09- 069 and the Spanish Consolider-Ingenio 2010
program CPAN CSD2007-00042.  TO wishes to thank M.M.~Fern\'andez for her
permanent support.


\end{document}